# Strained bubbles in van der Waals heterostructures as local emitters of photoluminescence with adjustable wavelength

Anastasia V. Tyurnina[1,4], Denis A. Bandurin[1], Ekaterina Khestanova[1], Vasyl G. Kravets[1], Maciej Koperski[1,2], Francisco Guinea[1,3], Alexander N. Grigorenko[1,2], Andre K. Geim[1,2], Irina V. Grigorieva[1,2]

[1]School of Physics and Astronomy, University of Manchester, Manchester M13 9PL, UK
[2]National Graphene Institute, University of Manchester, Manchester M13 9PL, UK
[3]IMDEA Nanociencia, Faraday, 9, Cantoblanco, 28049 Madrid, Spain
[4] Skolkovo Institute of Science and Technology, Nobel St 3, 143026 Moscow, Russia

***ABSTRACT****: The possibility to tailor photoluminescence (PL) of monolayer transition metal dichalcogenides (TMDCs) using external factors such as strain, doping and external environment is of significant interest for optoelectronic applications. Strain in particular can be exploited as a means to continuously vary the bandgap. Micrometer-scale strain gradients were proposed for creating 'artificial atoms' that can utilize the so-called exciton funneling effect and work, for example, as exciton condensers. Here we describe room-temperature PL emitters that naturally occur whenever monolayer TMDC is deposited on an atomically flat substrate. These are hydrocarbon-filled bubbles which provide predictable, localized PL from well-separated submicron areas. Their emission energy is determined by the built-in strain controlled only by the substrate material, such that both the maximum strain and the strain profile are universal for all bubbles on a given substrate, i.e., independent of the bubble size. We show that for bubbles formed by monolayer $MoS_2$, PL can be tuned between 1.72 to 1.81 eV by choosing bulk $PtSe_2$, $WS_2$, $MoS_2$ or graphite as a substrate and its intensity is strongly enhanced by the funneling effect. Strong substrate-dependent quenching of the PL in areas of good contact between $MoS_2$ and the substrate ensures localization of the luminescence to bubbles only; by employing optical reflectivity measurements we identify the mechanisms responsible for the quenching. Given the variety of available monolayer TMDCs and atomically flat substrates and the ease of creating such bubbles, our findings open a venue for making and studying the discussed light-emitting 'artificial atoms' that could be used in applications.*

Strain engineering is a powerful tool to tailor the physical properties of matter (1). Due to their remarkable stretchability, 2D materials - in contrast to their bulk counterparts - can withstand large strains before rupture, opening a new route to fabricate devices with strain-tunable electronic and optical properties. Many interesting effects were predicted (2,3) and observed by bending (4) and stretching (5,6) 2D materials. Among them, molybdenum disulphide ($MoS_2$) - a representative of transition metal dichalcogenides (TMDCs) - attracts a special attention from both fundamental and applied perspectives as it hosts a quality electronic system with a technologically relevant band gap (7-9). The large Young's modulus and high elastic strain limit (10) make 2D $MoS_2$ an important platform for strain engineering, too. Recent theoretical and experimental studies (11-19) have shown that the physical properties of $MoS_2$ are strongly affected by tensile or compressive strain. In particular, strain changes the bandgap as seen from, e.g., changes in the optical response (11,12,16-18), induced piezoresistivity (13) and direct-to-indirect gap transition (14,15).

One of the exciting opportunities presented by strain engineering using monolayer $MoS_2$ is the control of its PL. So far most studies employing strain as a means to engineer PL focused on creating point-like strain perturbations in monolayer TMDCs ($WS_2$, $WSe_2$) which act as quantum emitters (20-24). These are usually found at locations of sharp bending (21), wrinkles and folds (20), or sharply pointed nanobubbles (22-24). A different scenario can be realized using smooth gradients of biaxial strain as was proposed for monolayer $MoS_2$ in refs. (25,26): this creates a continuously varying bandgap profile that leads to a funneling effect such that photoexcited electron-hole pairs are driven towards areas of maximum strain where excitons recombine, producing locally a strongly enhanced PL. The resulting 'artificial atoms' should be able to absorb from a broad window of the solar spectrum and can be used as solar energy concentrators or as a basic element for exciton condensation and lasing (25). There have been significant efforts to realize such biaxial strains experimentally by, for example, depositing 2D $MoS_2$ on top of lithographically made nanostructures (27,28). This allowed relatively small strains of 0.2–0.5% and resulted in a shift of up to 50 meV for exciton emission and an enhanced PL intensity. Somewhat larger energy shifts were reported for suspended few-layer $MoS_2$ in ref. (28). The exciton funneling effect was also used to explain the PL shift in wrinkled few-layer $MoS_2$ (29). Nonetheless, the achieved strains and changes in the optical bandgap were relatively modest compared to the predictions (25) and difficult to control. In other experiments, strain of up to 7% was created by applying pressure to a suspended monolayer $MoS_2$ which resulted in large energy shifts but, unfortunately, the PL intensity was much reduced (11,18). The latter is consistent with the predicted (12,15,25) direct-to-indirect bandgap transition and a corresponding suppression of PL under uniform tensile strain (12). In the case of gradient strain, the PL suppression should dominate for membranes with areas >> 1 $\mu m^2$, as in refs. (11,12), which limits a practical use of the effect.

Here we propose an alternative approach where a smooth gradient strain of ∼ 2% can be reached within submicron areas of monolayer TMDCs. To this end, we have utilized hydrocarbon-filled bubbles that are naturally present in van der Waals (vdW) heterostructures created by depositing TMDC monolayers on atomically flat substrates (30). The bubbles are formed from contamination (adsorbed water and hydrocarbons) inevitably present on surfaces of assembled crystals (31). During the assembly, vdW forces attract the crystals together, which squeezes out trapped contamination into submicron-size pockets (bubbles) and leaves the remaining interface atomically sharp and free of contaminants. This process is usually referred to as self-cleansing (31,32) and typically occurs when monolayer graphene or TMDCs are deposited on lipophilic substrates (32). As shown in our earlier work (30), the shape of the bubbles and their aspect ratio $h/R$ (where $h$ is the maximum height and $R$ is the radius) are determined by just two characteristics of the vdW heterostructure: in-plane stiffness (Young's modulus $Y$) of the deposited monolayer and its adhesion energy $\gamma$ to the substrate. The theory yields (30) $h/R = \pi\gamma/5c_1 Y$ where the numerical coefficient $c_1 \approx 1$. Therefore, the aspect ratio for each 2D crystal/substrate combination is universal, independent of the bubble size. Furthermore, as the strain in the part of the monolayer enclosing a bubble is proportional to its aspect ratio, $\varepsilon \sim (h/R)^2$, the strain distribution is universal, too. Accordingly, one should expect universal PL characteristics for all the bubbles present on the substrate.

Combining monolayer $MoS_2$ with various atomically flat substrates, we demonstrate that the PL becomes highly localized spatially and its absorption/emission energies can be shifted by design of vdW heterostructures. In this approach, three complementary effects are utilized: (i) localization of strain within submicron-size bubbles so that the areas away from the bubbles remain unstrained; (ii) the fact that PL from the bubbles is not affected by a substrate and is similar to that from free-standing monolayer $MoS_2$ thanks to the spatial separation by trapped hydrocarbons (31,32); and (iii) strong quenching of PL from the surrounding areas that have good contact with (semi)conducting substrates. In particular, we show that PL from the bubbles is red-shifted in energy compared to unstrained flat regions by as much as 140 meV, with the shift determined by the adhesion to a particular substrate material. Also, the PL intensity notably increases from the least to the most strained bubbles. By quantifying the applied strain using atomic force microscopy (30) (AFM) we demonstrate a clear correlation of the observed PL shift and the increase in PL intensity with the built-in strain for different substrates.

*Sample preparation and characterization*

Bearing in mind that the strain in monolayers enclosing the bubbles depends not only on the monolayer crystal's elastic properties but also on its adhesion to a substrate (30), we fabricated a series of samples having different monolayer $MoS_2$/substrate combinations. To this end, $MoS_2$ flakes were deposited onto several atomically flat substrates: hexagonal boron nitride (hBN), monolayer- and few- layer graphene (further referred to as SLG and FLG, respectively), graphite, platinum diselenide ($PtSe_2$), tungsten disulphide ($WS_2$) and bulk $MoS_2$. Choosing different substrates allowed us to vary the adhesion; also only the substrates that ensure self-cleansing (see above) were chosen. To prepare the substrates, ~100 nm thick crystals were mechanically exfoliated from bulk materials and deposited onto an oxidized Si wafer. At this thickness their properties are the same as for bulk crystals but exfoliation allowed us to obtain clean, atomically flat surfaces. SLG and FLG were also deposited onto the $SiO_2$/Si wafer.

To make the heterostructures we have used the standard wet-transfer technique as described in detail elsewhere (32). Briefly, $MoS_2$ monolayers were first mechanically exfoliated onto a poly(methyl-methacrylate) membrane (PMMA). The latter was then loaded into a micromanipulator, placed face-down onto a chosen substrate and the PMMA was dissolved. The resulting heterostructures were heated (annealed) at 150°C for ~ 30 min, which resulted in spontaneous formation of a large number of sub-micron bubbles filled with hydrocarbons (31), with typical separations from ~ 1 μm to tens of microns (Fig. 1). The annealing time and temperature were optimized to ensure that the self-cleansing process was complete and the bubbles reached equilibrium conditions, that is, acquired their equilibrium shape and size (30). For the present study we selected only round or slightly elongated bubbles (Fig. 1c), as the aspect ratio-strain relation for bubbles of more complicated shapes – triangular, pyramidal etc. – is poorly defined (30).

We emphasize that the strains created by such bubbles (and their effect on PL discussed below) are qualitatively different from the much discussed 'nanobubbles' used to create quantum emitters in $WS_2$ and $WSe_2$ monolayers (20,22-24). In the latter case 'nanobubbles' are metastable, have irregular shapes and contain nano-folds or wrinkles that work as stress concentrators and point-like strain perturbations (33); as discussed by the authors (20,22-24) this is the reason they act as quantum emitters in a similar way to defects or edges. The bubbles in our work are in equilibrium and therefore stable and predictable, providing well-defined, smooth strain gradients that are universal for all bubbles on a particular substrate, i.e., pre-determined by the choice of monolayer-substrate combination and independent of the bubble size.

To correlate PL spectra with location and geometry of the bubbles, we first used AFM to obtain maps of bubble distribution and measure the topography of individual bubbles. An optical micrograph of a typical vdW heterostructure (MoS$_2$ on hBN) is shown in Fig. 1a and the corresponding AFM topography in Fig. 1b. The bubbles are seen as white spots on the AFM image (Fig. 1b) and the corresponding darkish areas on the optical micrograph (Fig. 1a). A zoomed-in AFM image of a typical bubble and its cross-sectional height profile are shown in Figs. 1c and d. For the substrates used in this work most of the bubbles had $R = 0.2 - 1$ µm and the aspect ratio $h/R$ varied from ≈ 0.11 to 0.15, resulting in different strains for different substrates. The largest $h/R \approx 0.15$ was found for MoS$_2$/graphite, MoS$_2$/hBN and MoS$_2$/MoS$_2$ heterostructures. Somewhat smaller $h/R$ of 0.14 and 0.13 were measured for WS$_2$ and SLG substrates, respectively, whereas bubbles with the smallest $h/R \approx 0.11$ and therefore the lowest strain were found in MoS$_2$ on PtSe$_2$.

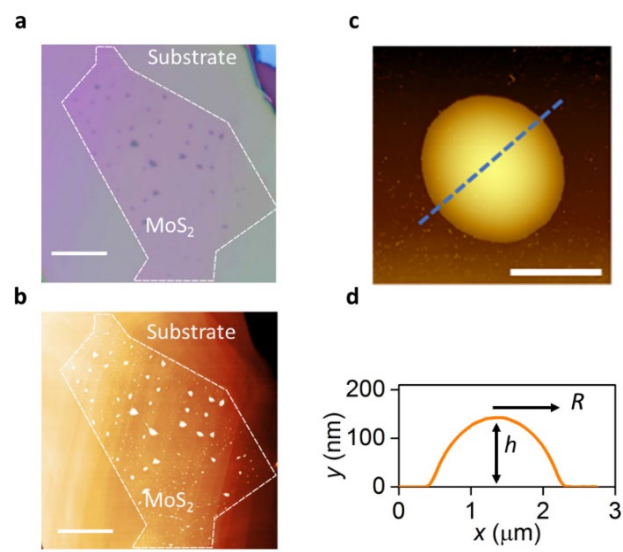

**Figure 1. MoS$_2$ bubbles.** (a) Optical micrograph and (b) AFM image of a MoS$_2$ flake transferred on top of an atomically flat crystal, in this case ≈ 260 nm thick hBN. White dashed line shows the border of the MoS$_2$ monolayer. Scale bar, 10 µm. (c) AFM image and (d) cross-sectional profile of a bubble formed by a MoS$_2$ monolayer on a graphite substrate. Scale bar 2 µm. Dashed blue line in (c) indicates where the profile in (d) was measured.

PL spectra were measured in the backscattering configuration using Horiba Raman system XploRA$^{TM}$ PLUS with the spectrometer grating of 600 grooves·mm$^{-1}$, which corresponds to the spectral resolution of about 5 cm$^{-1}$ (0.003 meV). The excitation laser energy was 2.33 eV and the spot size ≈ 1 µm. All spectra were collected at room temperature in the energy range between 1.2 and 2.3 eV at laser power 0.011 and 0.125 mW. There was no dependence of the PL spectra on the laser power in this range, i.e. all measurements were done in the linear regime. The spectra were collected using the same conditions (acquisition time and focal distance) from three different areas: on top of a bubble, on the flat area near the bubble (no, or very little, strain) and, for reference, from the substrate area not covered by a MoS$_2$ monolayer (Fig. 2b). The laser spot size was smaller than the diameter of our larger bubbles, such as that shown in Fig. 1c. These were typically found on graphite and MoS$_2$ substrates. For hBN, SLG and PtSe$_2$ substrates all bubbles were smaller than the spot size, so that the spectrum taken from a bubble unavoidably contained a contribution from the surrounding flat area (also see below). We note that it would simplify comparison if bubbles of similar sizes could be found on different substrates, but unfortunately it is not possible: As was shown in our earlier study (30), each monolayer-substrate combination tends to have a certain range of sizes, with the substrate being mostly responsible for the difference. For example, $MoS_2$ bubbles on bulk $MoS_2$ tended to have larger sizes than $MoS_2$ bubbles on hBN, bubbles formed by SLG on hBN substrate were significantly larger than those formed by monolayer hBN and so on. These are empirical findings explaining that for some substrates ($PtSe_2$, SLG) we could not find bubbles larger than the laser spot size.

*Experimental results*

A dramatic effect of the bubbles on $MoS_2$ photoluminescence is clear from Fig. 2a which compares the PL spectra taken from a bubble, where $MoS_2$ is strained and separated from the substrate by a layer of hydrocarbons, and from the flat, unstrained part of the same heterostructure, where the $MoS_2$ monolayer and graphite substrate are in immediate contact. In the flat areas, PL is strongly suppressed, with a peak intensity of just 4% of that from the bubble area. On the other hand, the emission energy from the flat area, seen as the peak centered at around 1.88 eV, is practically the same as for free-standing or unstrained $MoS_2$ in earlier studies (8). In the literature this peak is usually associated with radiative recombination of neutral excitons (8) and/or charged trions (34). A much smaller peak at higher energies probably involves excitonic transitions between the conduction band and the deeper spin-split energy levels in the valence band (8). The PL spectrum from the bubble is very different: First, the main PL peak is strongly shifted to lower energies, by $\Delta E \approx 120$ meV, and, second, its integrated intensity is over 50 times greater than for the flat area (left inset to Fig. 2a). The ratio of integrated intensities in this case is about 55, while for $MoS_2$ and $WS_2$ used as substrates the ratio is much larger, exceeding 1000.

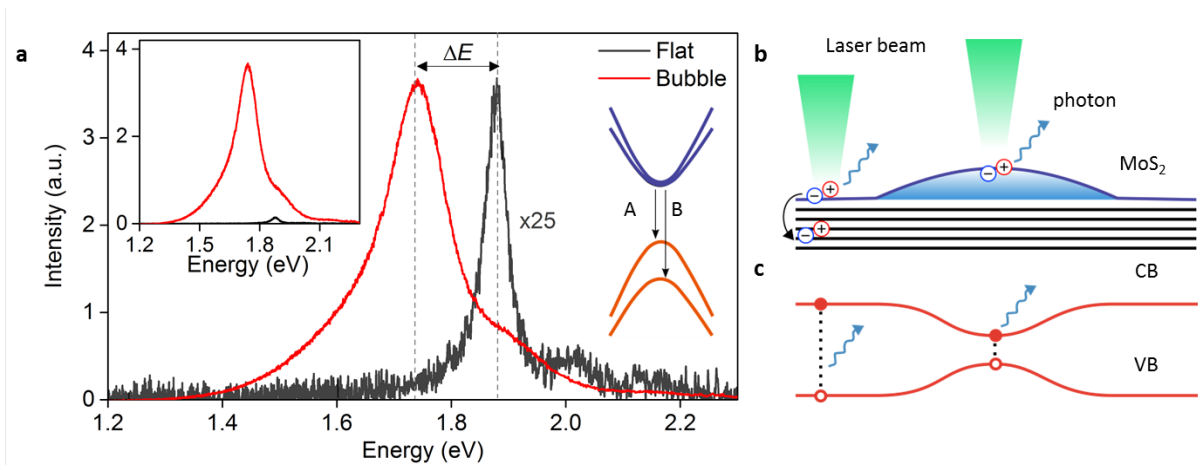

**Figure 2. Photoluminescence from $MoS_2$ bubbles.** (a) PL spectra from the bubble (red) and from the flat region (black) measured at room temperature. The substrate is 44-nm-thick graphite. For comparison, the spectral intensity from the flat region is normalized to that of the bubble. Left inset: as-measured (not-normalized) PL spectra. Right inset: Schematics of $MoS_2$ band structure. (b) Sketch of the bubble formed between an atomically thin film and a substrate. Green beams depict laser light focused on a flat and curved parts of $MoS_2$ (not to scale). (c) Band diagram profile showing a decrease of the band gap towards the bubble center. CB and VB denote conduction band minimum and valence band maximum.

Fig. 3 compares PL spectra from the bubbles and flat areas for all the different substrates used in our study. To compare PL intensities, the spectrum from the bubbles for each $MoS_2$/substrate combination is normalized to its peak intensity and the corresponding flat-area spectrum multiplied by the same factor. It is clear that the PL signal acquired from the flat regions shows large variations vs the substrate material. Their PL response broadly falls into three groups: The signal with the strongest intensity was detected from the flat areas of MoS$_2$ on top of hBN and on monolayer graphene (SLG). Here the spectra from the bubbles and flat areas are rather similar, the only difference being a small shoulder at low energies visible on the bubble's spectrum for MoS$_2$/SLG. It requires subtraction between the normalized spectra for unstrained and strained regions to separate a tiny peak at 1.78 eV coming from the bubbles on the SLG substrate. As a result the spectrum on SLG appears to be comprised of three peaks: two peaks corresponding to A and B excitons originating from the flat, unstrained part of the MoS$_2$ monolayer (see Fig. 2a), and the low-energy shoulder due to the bubble. For hBN the PL signal from flat areas is even stronger, completely dominating the overall PL (the bubbles in MoS$_2$ on hBN are much smaller than the laser spot size, 200-300 nm in diameter, and contribute less than 10% to the overall PL). In this case it was not possible to separate the signal from the bubbles with a sufficient accuracy. For the PtSe$_2$ and graphite substrates, PL from flat areas is strongly quenched, with much higher intensity from the bubbles compared to the flat areas. Finally, for bulk MoS$_2$ and WS$_2$ the flat-area PL is reduced in intensity by more than two orders of magnitude;

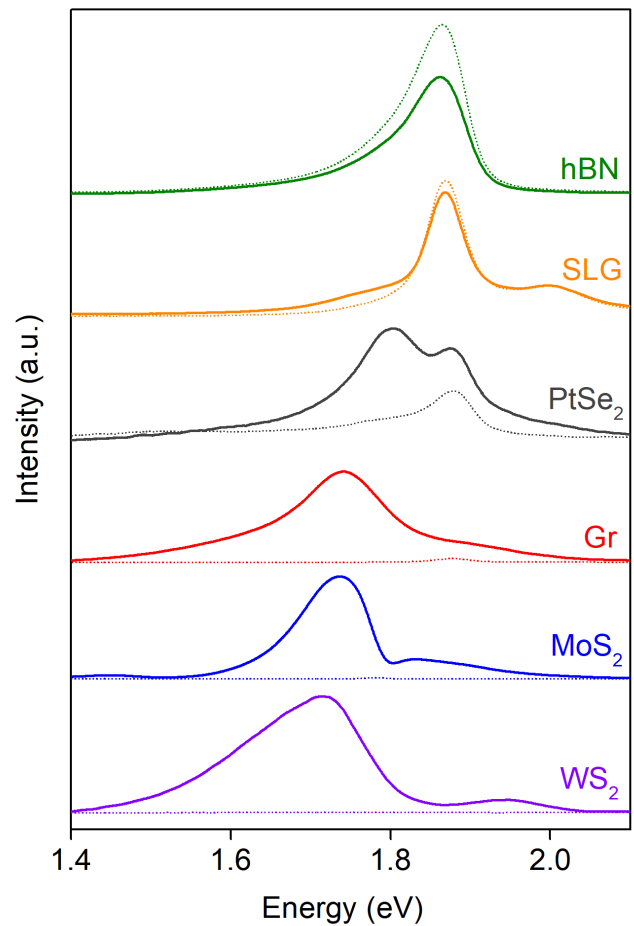

**Figure 3. PL from monolayer MoS$_2$ on different substrates.** Thick solid lines show PL measured from strained bubbles and dashed lines show PL from the flat areas. To allow comparison, for each substrate the spectrum from the bubbles is normalized to its peak intensity and the corresponding flat-area spectrum is multiplied by the same factor. Spectra for different substrates are shifted for clarity. The two-peak structure for PtSe$_2$ is due to the small size of all bubbles in this case, i.e., the higher-energy peak originates from the surrounding flat areas (see text).

for WS$_2$ substrate PL is almost fully quenched (at the measurements noise level). Where visible (for hBN, SLG, graphite and PtSe$_2$ substrates) the PL peak position for the flat areas shown by dashed lines in Fig. 3 is almost constant (varies within 0.5 %). In contrast, the PL energies for the bubbles differ significantly from those of unstrained MoS$_2$.

The strong suppression of PL from the areas of good contact between monolayer MoS$_2$ and the substrates in our heterostructures is in principle not surprising and is in agreement with literature where large variations of PL intensity were reported for TMDC monolayers deposited on different substrates (35-44). For example, a factor of 10 suppression of PL intensity was reported for monolayer MoS$_2$ on 15nm-thick graphite, while on mica, a polymeric film or hBN it was affected very little (36); PL from monolayer WSe$_2$ on SiO$_2$ showed a non-monotonic dependence on SiO$_2$ thickness, with enhanced PL intensity for certain thicknesses (41), and so on. These effects have been attributed predominantly to multiple reflections from interfaces between the constituent layers of a heterostructure and the resulting interference, as well as to doping, charge transfer and defect-assisted non-radiative recombination of electron-hole pairs (36,37,41-44). For each TMDC/substrate combination, the effect on PL intensity was shown to be highly sensitive to the substrate material, layer thickness and composition, optical constants, presence of defects and/or charge-donating molecules (35-44). Specific mechanisms responsible for the PL quenching in our heterostructures (Fig. 3) will be discussed below. Now we focus on PL from the bubbles and its relation to the built-in strain.

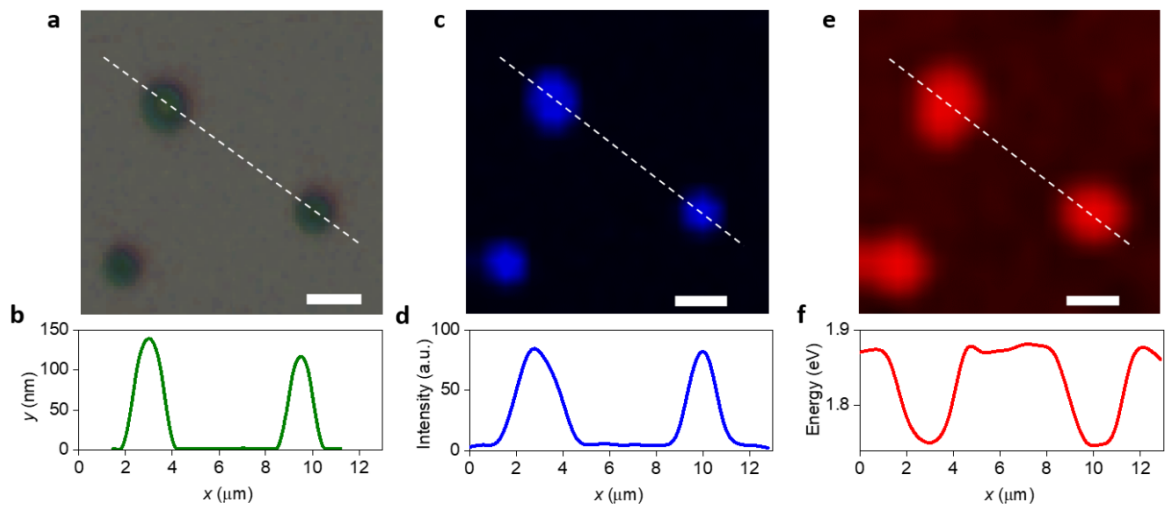

**Figure 4. Lighting up MoS$_2$ bubbles**. (a-b) Optical photograph (top) of three bubbles formed between MoS$_2$ monolayer and a graphite substrate (a) and corresponding AFM profiles (b) measured along the white dashed line. (c-d) Map of the PL intensity (c) integrated over the spectral range $1.4 - 1.95$ eV and the corresponding profile (d) along the dashed line. (e) Map of the PL peak energy. The dark color corresponds to the peak of unstrained MoS$_2$ at $E = 1.88$ eV whereas red is for $E = 1.74$ eV. (f) Profile of the PL peak position along the white dashed line. Scale bar, 2 μm. $\lambda = 532$ nm.

To quantify the different PL responses as a function of the substrate, we identified PL peak positions using standard Gauss-Lorentz fitting in LabSpec6 software. For the substrates where PL from flat areas was not strongly quenched (PtSe$_2$, SLG) we were able to disentangle the PL signal coming from the bubbles by taking the difference between the two spectra, such as shown in Fig. 3. This allowed us to identify the peak PL energies for the bubbles on all different substrates.

To compare the PL from the bubbles found on the same heterostructure but having different diameters and heights, Fig. 4 shows the PL energy and intensity mapped onto optical AFM images for the MoS$_2$/graphite stack. All bubbles are clearly seen at the same positions on all three maps. As the PL intensity from the bubbles is much higher than from the flat areas and the emission energy significantly lower, the bubbles appear as bright spots on both energy and intensity maps. In fact, the profiles of PL intensity follow very closely the cross-sectional profiles obtained by AFM, as shown in Figs. 4b and d. For small bubbles, with diameters smaller than the laser spot size, we found that the PL intensity depended on the ratio between the bubble diameter and the spot size. This is to be expected because a part of the PL response is collected from the surrounding flat areas where the PL intensity is strongly reduced.

In contrast to the PL intensity, which varied from bubble to bubble due to their different sizes, the shift of the PL *energy* at the center of each bubble, $\Delta E$, was universal, i.e. found to be the same (within a few percent) for all bubbles in the same heterostructure (Fig. 4f and Supplementary Figure S1). This is in agreement with the universal value of the strain associated with bubbles for a given monolayer/substrate combination (30). Furthermore, in agreement with the strain distribution across a bubble (which is maximum at the top), the largest energy shift is observed in the central area of the bubble as shown in Fig. 4f. Fig. 5a plots the shift of the bubbles' PL energy, $\Delta E$, relative to the peak position for MoS$_2$ monolayer on hBN. The latter case is used as a reference because its PL peak position is practically unaffected either by the substrate or by the presence of the bubbles. The smallest shift $\Delta E \approx 50$ meV was found for the bubbles formed by monolayer MoS$_2$ on PtSe$_2$, whereas the strongest ones, $\Delta E \approx 120 - 140$ meV, were for MoS$_2$/graphite, MoS$_2$/MoS$_2$ and MoS$_2$/WS$_2$ heterostructures, in agreement with the estimated strain values (respective aspect ratios $(h/R)^2$) – see Fig. 5a.

To compare the PL *intensity* for bubbles on different substrates, it was necessary to exclude the contribution from the surrounding flat areas for small bubbles (see above). To this end, for each spectrum from a bubble that was smaller than 1 μm in diameter (laser spot size), we first subtracted the measured background PL from the flat part of the heterostructure and then normalised the spectral intensity to the area of the laser spot, taking into account the bubble's area measured on AFM topography maps (as in Fig. 1) and the intensity distribution across the bubble (such as in Fig. 4d). The results are shown in Fig. 5b,c where Fig. 5b compares peak PL intensities measured from the bubbles (red symbols) and flat areas (blue) for different MoS$_2$/substrate combinations and Fig. 5c shows the evolution of the integrated PL intensity with the aspect ratio of the bubbles, i.e. with strain. While the PL intensity from areas of good contact between the monolayer MoS$_2$ and the substrate varies by orders of magnitude (note the logarithmic scale in Fig. 5b), the peak PL intensity for the bubbles on different substrates varies relatively little, by about 50%. Nevertheless, these variations are clearly not random but are correlated with the value of strain: the lower values of peak PL intensity correspond to less strained bubbles (on PtSe$_2$, SLG, and WS$_2$ substrates). This trend is much clearer in Fig. 5c that shows the integrated PL intensity vs $(h/R)^2$ (i.e., maximum strain) on a linear scale.

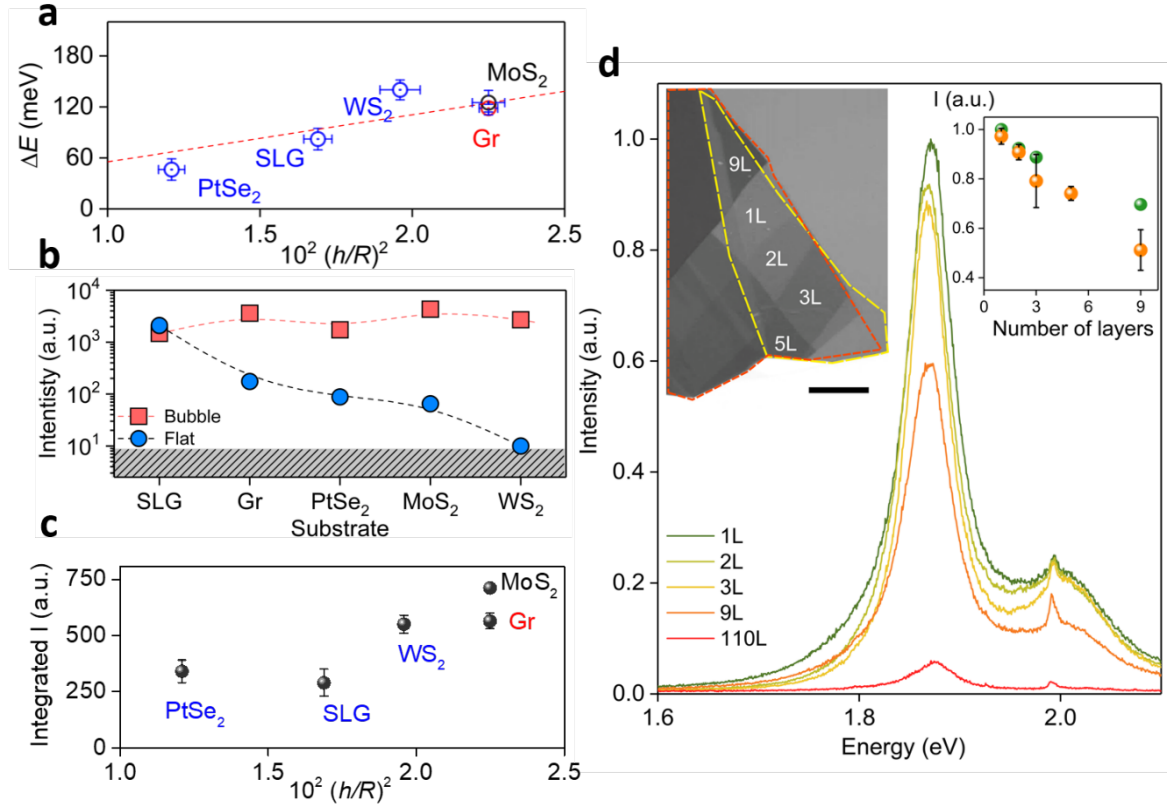

**Figure 5. Variable photoluminescence of monolayer MoS$_2$.** (a) Shift, $\Delta E$, of the PL peak from strained MoS$_2$ bubbles as a function of $(h/R)^2$ for different substrates. $\Delta E$ for each substrate (symbols) is determined relative to the PL peak of the unstrained MoS$_2$ deposited onto hBN. Red dashed line is the best linear fit to the experimental data. Error bars represent the variation of PL energy for all studied bubbles within a sample (see, e.g., Fig. S1) and the accuracy of $(h/R)^2$ measurements. (b) PL peak intensity measured from the flat regions (blue circles) and from the bubbles (red squares) for MoS$_2$ monolayers deposited onto different substrates. Dashed lines are guides to the eye. The grey dashed area indicates the noise level in our PL measurements. (c) Integrated PL intensity from the bubbles as a function of $(h/R)^2$. (d) PL spectra of MoS$_2$ monolayer deposited onto a FLG substrate with several steps corresponding to different numbers of graphene layers, *N*. Left inset: Optical image of the MoS$_2$/FLG stack. Yellow and red dashed lines show the boundaries of MoS$_2$ and FLG flakes, respectively. The number of graphene layers for each step is indicated by labels. The sharp peak around 2 eV is a Raman mode of graphene. Right inset: Orange dots show the measured peak PL intensity vs *N*, normalized to its value for *N* = 1 (data for *N* = 110 is not shown as the intensity there is suppressed by a factor of 25). Green dots show the corresponding results for the peak PL intensity calculated from measured optical reflection coefficients vs *N* using Fresnel theory (see text).

*Discussion: variable emission wavelength and funneling effect.*

To understand the observed strong red-shift of the PL energy for the bubbles we recall that the band gap of atomically thin TMDCs, including monolayer MoS$_2$, is strongly modified by strain (14,15,45) while the exciton binding energy remains unaffected (25,35), which results in a shift of the exciton emission energy. In the case of bubbles, the strain is biaxial and strongly localized in the area where the monolayer is lifted away from the substrate (cf. AFM topography and PL locations in Fig. 4). The maximum strain is found around the center of a bubble, creating a gradient towards its edges, approximately replicating the bubble's cross-sectional profile (30). As both the maximum strain and the profile are universal (equal) for all the bubbles within the sample (30), the decrease in the peak PL energy from the bubbles relative to the unstrained monolayer $MoS_2$ should be equal, too, in agreement with our observations. From the fit of $\Delta E$ vs $\varepsilon \sim (h/R)^2$ in Fig. 5a we obtained an energy shift of 55 meV per 1% change in the maximum strain, in good agreement with literature values for monolayer $MoS_2$ (11,12,16,17).

The PL intensity also clearly correlates with strain. This is noticeable for the peak PL intensity (Fig. 5b) and is particularly clear for the integrated intensity that at least doubles as the maximum strain increases from $MoS_2$ on $PtSe_2$ to $MoS_2$ on bulk $MoS_2$ and graphite substrates (Fig. 5c). This is in contrast to earlier observations on large ($\gg 1\ \mu m^2$) strained $MoS_2$ membranes (11,12,18), where the PL intensity was strongly suppressed with increasing strain [by a factor 2-3 for a similar range of strains as in our experiments (11,12)]. We attribute this result to two opposing effects that are simultaneously present in our ~1μm diameter bubbles and partially compensate each other: Suppression of the PL intensity due to direct-indirect optical bandgap transition under tensile strain (12,25) and its increase due to the funneling effect (25,26). In the latter case, as the maximum strain is localized at the top of a bubble, tapering off towards its base, it produces gradual bending of the electronic bands towards the minimum band gap at the top/center of the bubble, leading to a potential gradient for the charge carriers (25-27) – see Fig. 2c. The latter then acts as a driving force for photo-generated excitons that are attracted to the potential valleys at the bubble center, enhancing the PL. The energy gradient seen by excitons and, therefore, the PL enhancement has been predicted to increase in proportion to the strain gradient, $\nabla \varepsilon$ (25,26), which in the case of bubbles is proportional to the maximum strain (30). It is important to note that, for the funneling effect to dominate, the excitons generated at different points of the strain gradient must remain bound while being 'funneled' towards the maximum strain, i.e. the drift length $l_{\text{drift}}$ before recombination should exceed the bubble's radius (25). Similarly to the spatial variation of the exciton energy (the funneling driving force), the drift length is proportional to the strain gradient, too (25), $l_{\text{drift}} \propto \nabla \varepsilon$. In an ideal $MoS_2$ monolayer and for large gradients, $\nabla \varepsilon \sim 10\%/\mu m$, $l_{\text{drift}}$ was estimated to be ~600 nm (25). In real monolayers and for smaller $\nabla \varepsilon$, as in our case, this distance is likely to be shorter, falling well below the radius of most our bubbles, which would reduce the peak PL intensity. This can explain our observation that, while the peak PL intensity clearly increases from the least ($MoS_2$ on $PtSe_2$) to the most ($MoS_2$ on graphite, $MoS_2$ on bulk $MoS_2$) strained bubbles, this increase is relatively modest. It can also be expected to broaden the PL peaks due to radiative recombination of excitons in all areas of the bubble (rather than just at the top) where the strain and therefore the optical bandgap are different. Such broadening and a corresponding much larger increase in integrated intensity are indeed observed in our experiments.

*Quenching of the PL by the substrates.*

As demonstrated in Fig. 3 and 5b, the PL signal acquired from the flat (unstrained) regions of our heterostructures strongly varied with the substrate material, with particularly strong PL quenching observed for graphite, bulk $MoS_2$ and $WS_2$ substrates. In principle, a strong effect of the substrate on the PL intensity is to be expected but the exact mechanism underlying such quenching depends sensitively on the details of a heterostructure (35-44). In our case understanding the quenching mechanism is particularly important as it ensures the possibility to localize PL on the bubbles only.

In literature several different mechanisms responsible for PL suppression and/or changes of the emission energy have been identified when a TMDC monolayer is in proximity to a metallic or a semiconducting surface: (i) dielectric screening that reduces the exciton binding energy (38); (ii) charge and/or energy transfer that facilitates non-radiative recombination of electron-hole pairs and reduces the exciton lifetime (37,39,46-48); (iii) optical interference between incident light waves and those reflected from interfaces (36,41,46). Screening appears to play only a minor role in our experiments as the position of the main exciton peak for $MoS_2$ on hBN, SLG, graphite and $PtSe_2$ is practically unaffected (Fig. 3). On the other hand, both the charge transfer and modification of local electromagnetic fields and local density of optical states caused by interference in the multi-layered structure can be expected to be significant (36,37,41-44,46). In the case of semiconducting $MoS_2$ or $WS_2$ substrates, an additional effect of band alignment can lead to the transfer of charge to the substrate and non-radiative relaxation (37,39,40) which was seen, for example, in heterobilayers (39,40,43) (stacked together monolayers of different TMDCs).

To check whether optical interference indeed plays a leading role in our observations, we have conducted additional experiments, depositing a $MoS_2$ monolayer on a terraced few-layer graphene (FLG) crystal, where different parts of the structure correspond to different numbers of graphene layers, *N*, and measuring its PL and optical reflection coefficient. The sample is shown in the left inset of Fig. 5d where a $MoS_2$ monolayer (yellow dashed line) is deposited on a FLG crystal (orange dashed line) with the step heights corresponding to one or more graphene layers. The corresponding PL spectra taken from regions with different *N* are shown in the main panel of Fig. 5d and the wavelength-dependent optical reflection coefficient vs *N* is shown in Supplementary Fig. S2.

The PL peak intensity decreases sharply as a function of $N$ (right inset in Fig. 5d, orange dots) while the reflection of visible-to-infrared light shows a deep Fabry-Perot resonance (reflection minimum, Supplementary Fig. S2) at wavelengths corresponding to the main excitonic emission for the $MoS_2$ monolayer, $600 - 660$ nm (Fig. 5d). It is well known that the PL intensity is affected very strongly by light interference in the vicinity of a Fabry-Perot resonance, which explains the strong quenching of PL by just a few graphene layers seen in our experiment. Furthermore, the amount of reflected light at the PL peak energy, 1.88 eV, notably depends on the number of graphene layers *N,* which implies a similar *N* dependence for the PL intensity. Using the reflection data in Supplementary Fig. S2a and Woollam's WVASE32 software package for spectroscopic ellipsometry, we have calculated the corresponding PL suppression factor arising from the change in the local electric field and in the emission pattern due to the interference of PL photons emitted directly into open space and reflected from the interfaces in $FLG/SiO_2/Si$ stacks (46). The result for the full energy range is shown in Supplementary Fig. S3. The right inset of Fig. 5d compares the calculated suppression factor at 1.88 eV and the observed reduction in peak PL intensity vs *N*. There is a good agreement with the overall trend seen in our PL measurements, indicating that modification of the local optical field is indeed the main factor responsible for the observed, remarkably strong, PL suppression. The somewhat larger PL suppression found experimentally for larger *N* (see right inset in Fig. 5d) is probably an indication that charge/energy transfer also start playing a role for thicker FLG/graphite substrates.

A similar agreement between the measured and estimated PL quenching due to the modification of the optical field was found for bulk $MoS_2$ and $WS_2$ substrates. Using the optical constants, $n \approx 5$ and $k \approx 1$ (see Supplementary Information for details) we obtained a PL suppression factor for these substrates $S \approx 0.012$. This is close to the measured ratio of PL intensities from the bubbles and flat areas of $MoS_2/(bulk)MoS_2$ stacks, $\sim 0.01$ (see Fig. 5b), but significantly less than the PL suppression for MoS$_2$ on bulk WS$_2$ (measured $S\sim10^{-4}$). The latter is an indication that modification of the optical field is only partially responsible for the PL quenching by WS$_2$ substrate, with a bigger role played by non-radiative transfer of energy from the light-induced excitons to either single-particle or collective excitations in the substrate itself (e.g. excitons, phonons). The fact that bulk WS$_2$ has an excitonic peak at 2 eV (Supplementary Fig. S3) provides a natural explanation for the bigger role of non-radiative energy transfer in this case.

While modification of the optical field in principle also applies to the bubbles, the mirror-image dipole in this case is separated from the initial PL dipole by the double height of the bubble, $2h\sim100-200$ nm, which reduces the suppression factor to $\sim1$ and removes the possibility of non-radiative energy transfer completely. The above analysis demonstrates an effective route to localize photoluminescence in the selected (sub)micron-size bubbles, for example, by ensuring vanishing reflection from the substrate or choosing a substrate with known single-particle or collective excitations at energies corresponding to the excitons in the TMDC monolayer.

Finally we note that, in addition to PL quenching, comparison of the spectra in Fig. 2 shows that the asymmetry of the PL peaks from the bubbles also depends on the type of substrate: PL peaks are asymmetric for semiconducting WS$_2$ and MoS$_2$ substrates, but the asymmetry is absent for MoS$_2$ bubbles on graphite. We speculate that this can be due to the effect of other excitations of the system at lower energies. In particular, interlayer excitons (50) are likely to exist near the edges of the bubbles where the monolayer MoS$_2$ is very close to the substrate but is still strained and the PL not fully quenched. The energy of these excitons depends on the relative alignment of the gap edges in the two layers; typically they should be found at lower energies than the intralayer excitons, contributing to the low-energy 'tail' on the spectra.

In conclusion, we demonstrate that a ubiquitous, easily accessible, and usually avoided feature of van der Waals heterostructures (hydrocarbon-filled submicron-size bubbles formed by TMDC monolayers on atomically flat substrates) provides ideal conditions for realization of localized PL with reproducible characteristics: large shift of the PL emission energy, pre-determined by the choice of the substrate material, and a significant enhancement of the PL intensity due to exciton funneling. Using bubbles in monolayer MoS$_2$ transferred onto $\sim100$ nm thick MoS$_2$, WS$_2$, PtSe$_2$ and graphite substrates we have demonstrated that the peak PL energy can be varied from $E\approx1.81$ eV to 1.72 eV, with the same $E$ for all bubbles in each studied heterostructure. This PL energy shift is notably larger than could be achieved so far for MoS$_2$ monolayers without a reduction in PL intensity, as the funneling effect in our experiments ensures that the PL intensity is not only undiminished for strains as large as 2%, but shows a notable increase. In addition, our study sheds light on the mechanisms of PL quenching of MoS$_2$ monolayers transferred onto different 2D materials and provides a 'recipe' to ensure strong PL quenching everywhere except for the regions of localized strain (bubbles). Our findings open a new route to create the proposed light-emitting 'artificial atoms' that can be used in a variety of applications, for example, as unique identifiers for cryptographic key generation: Since the bubble positions are unique for every sample while all of them are characterized by the same PL emission energy and are the only objects visible on PL intensity maps (as we show in Fig. 4c), they can be used as optical identifiers following a similar scheme to the recent proposal in ref. (49).

# SUPPLEMENTARY INFORMATION

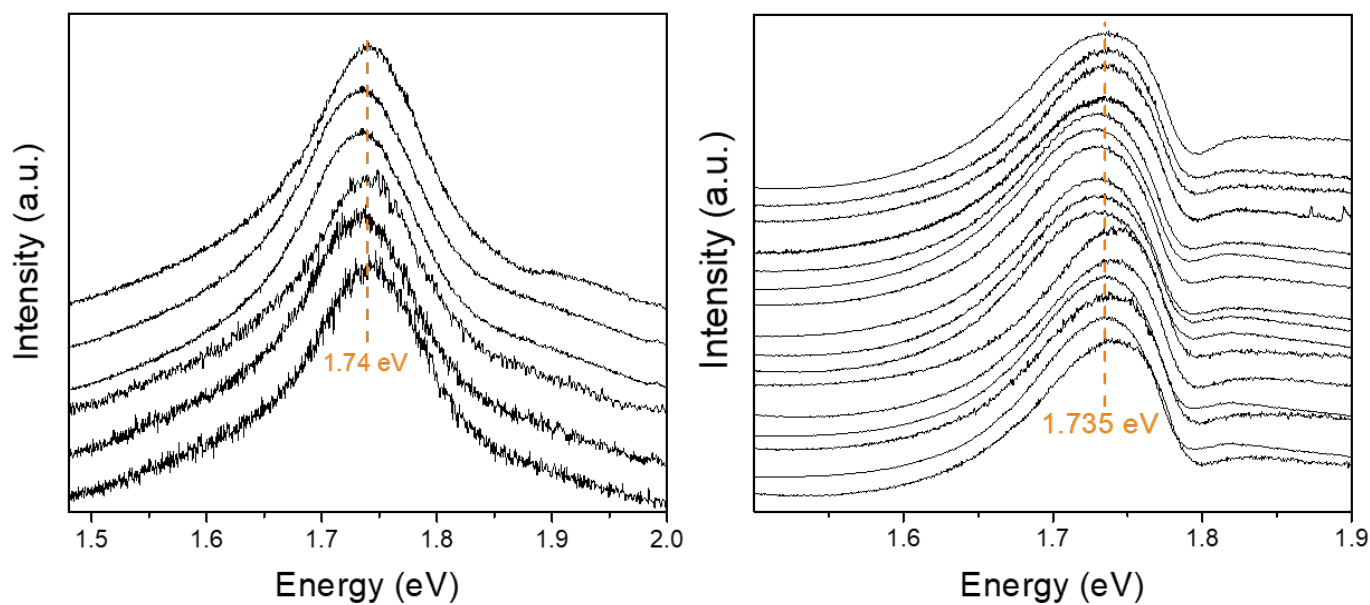

**Figure S1. Reproducibility of PL spectra from different bubbles on the same substrate and PL stability over time.** PL spectra acquired from MoS$_2$ bubbles of different sizes on the same bulk graphite substrate (left) and on the same bulk MoS$_2$ substrate (right). Each spectrum is normalized to its maximum intensity; the spectra are shifted for clarity. All PL measurements were done in the backscattering geometry, under x100 optical magnification using the same excitation energy of 2.33 eV but at different power (0.011 or 0.125 mW) and using different acquisition times; this explains the higher level of noise for some of the spectra compared to others. Importantly, both power levels used in the measurements corresponded to the linear regime, i.e., were well below the threshold power that can cause heating of the bubbles. This was verified by preliminary power-dependent measurements, confirming the absence of heating effects for powers below 0.15 mW. Several spectra for the MoS$_2$ substrate (write panel) correspond to repeat measurements on the same bubbles shortly after the sample was prepared and after 12 months. One spectrum from the right figure (4$^{th}$ from the top) was acquired at a different laser excitation, 1.96 eV; as expected in this case, sharp Raman peaks are visible between 1.85 and 1.9 eV, arising from MoS$_2$ and Si substrates, respectively.

To measure the wavelength-dependent reflection coefficient and evaluate the expected PL suppression factor for our MoS$_2$/FLG/SiO$_2$/Si stacks as a function of the number of graphene layers *N*, we used an ad-hoc spectrometer with focusing optics in the reflection mode. The measurements were done on the same sample as that used for photoluminescence experiments where a MoS$_2$ monolayer was transferred onto a 'staircase' few-layer graphene (FLG) crystal (see Fig. 5c in the main text). The incident light from the laser driver light source (LDLS) was focused on the sample with the FL 40x objective and then collected by the same objective. The reflected light was focused on the entrance of an optical fiber (200 μm core) coupled to the Ocean Optics USB2000 spectrometer. The reflection spectra were recorded by normalizing the raw spectra measured from the sample to the spectra measured from the bare SiO$_2$/Si substrate. For comparison, the *N*-dependent reflection coefficients for MoS$_2$/FLG/SiO$_2$/Si stacks with layer thicknesses corresponding to the experiment (see main text) were modelled using the transfer-matrix approach and Fresnel equations (1,2). Figure S2 shows both the measured and calculated reflection coefficients, with a good agreement between the two.

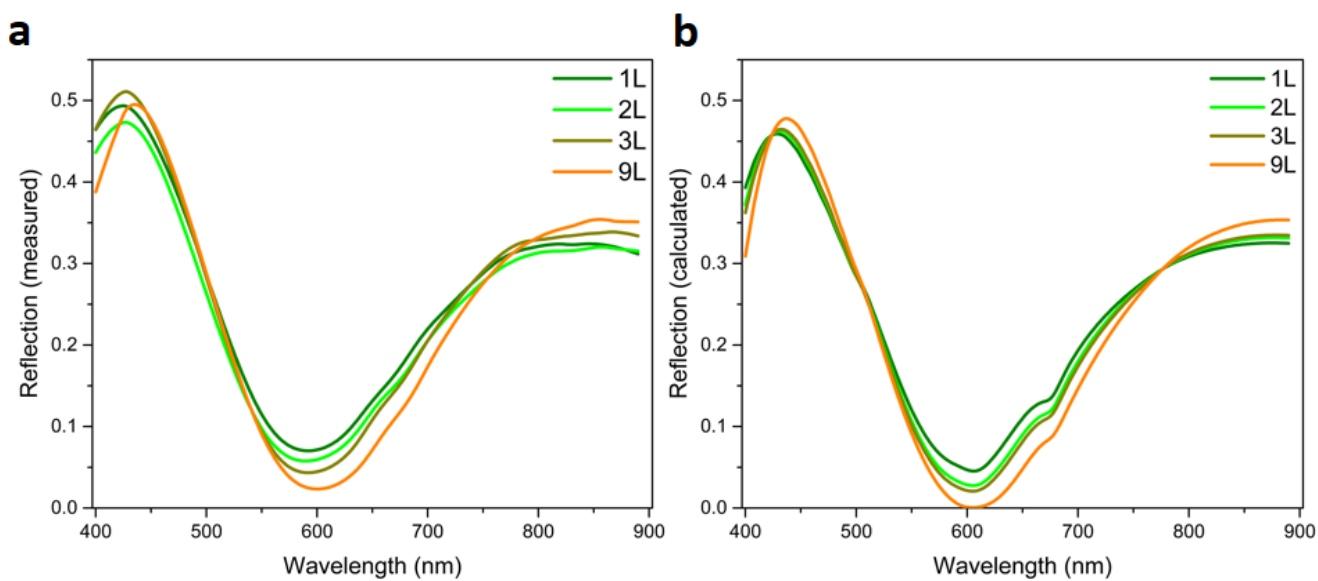

**Figure S2. Wavelength-dependent reflection coefficients for the MoS$_2$ monolayer deposited on a composite FLG/SiO$_2$/Si substrate with different numbers of graphene layers, *N*. (a)** Measured absolute reflection coefficients for normal incidence. Raw data have been smoothed and normalised to the reflection from the bare SiO$_2$/Si substrate. Different colours correspond to different *N* (see legends). **(b)** Calculated reflection coefficients for the studied heterostructure (monolayer MoS$_2$ on the substrate composed of *N* layers of graphene/290 nm SiO$_2$/bulk Si).

Using the reflection data in Figure S2a and Woollam's WVASE32 software package for spectroscopic ellipsometry, we have calculated the corresponding PL suppression factor arising from (i) the change in the local electric field (produced by the incident wave) in the MoS$_2$ monolayer caused by light interference within the MoS$_2$/FLG/SiO$_2$/Si stacks and (ii) changes of the emission pattern caused by interference of PL photons emitted directly and reflected from the interfaces in FLG/SiO$_2$/Si stacks (see main text). The corresponding evolution with *N* of the enhancement/suppression factor for PL intensity as a function of wavelength is shown in Figure S3.

To evaluate the expected PL suppression for monolayer MoS$_2$ in perfect contact with bulk MoS$_2$ or WS$_2$ substrates (in order to compare with the PL data for flat parts of the studied heterostructures) we have extracted the spectral dependences of the complex refractive index $\tilde{n} = n + ik$ for mechanically exfoliated ~100nm thick crystals of MoS$_2$ and WS$_2$ using spectroscopic ellipsometry. The ellipsometric measurements were performed with a J.A. Woollam VASE variable angle ellipsometer (M-2000F) with a 30 μm focal spot in the wavelengths range of 240–1700 nm. The ellipsometric data were modelled using WVASE32 software based on Fresnel coefficients for multilayered films. The extracted optical constants $n$ and $k$ are shown in Figure S4. At the relevant PL energy of ~1.9 eV (corresponding to the main excitonic peak of an unstrained MoS$_2$ monolayer) the extracted optical constants for bulk MoS$_2$ and WS$_2$ are very similar, $n \approx 5$ and $k \approx 1$, yielding an overall PL suppression factor calculated with our software package as $S \approx 0.012$.

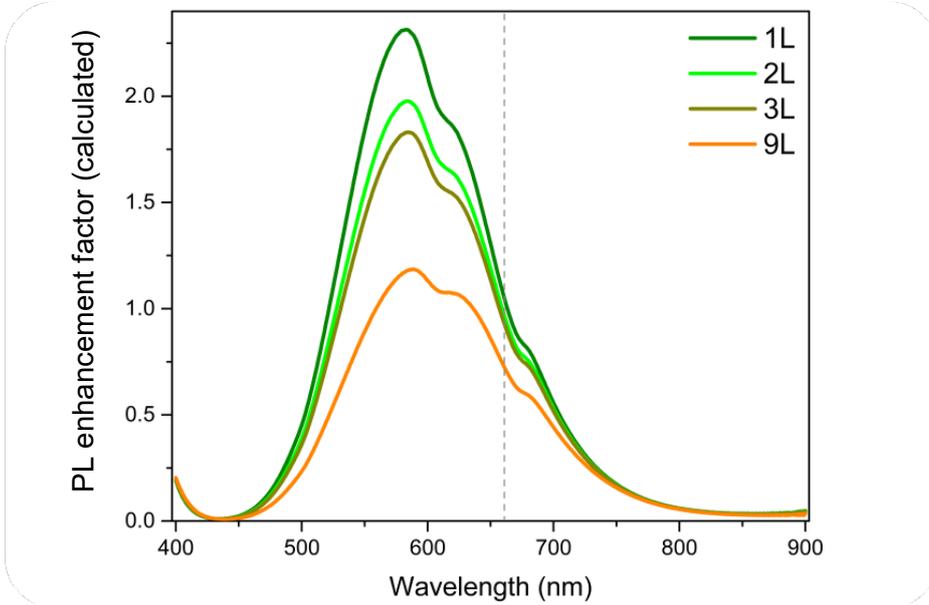

**Figure S3. Calculated wavelength-dependent enhancement/suppression factor of fluorescence intensity for MoS$_2$ monolayer on a composite FLG/SiO$_2$/Si substrate with different numbers of graphene layers *N*.** Different colour curves correspond to different *N* (see legend). Dashed line indicates the wavelength (660nm) corresponding to the peak PL intensity observed in experiment (Fig. 5c). The corresponding intensity for each *N* is shown in the right inset of Fig. 5c in the main text by green circles.

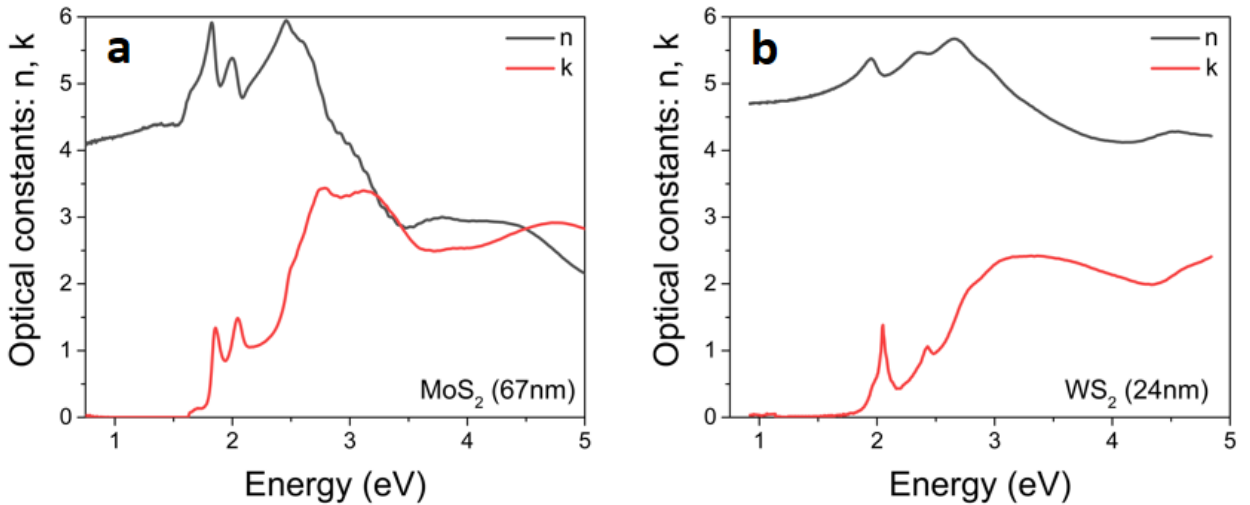

**Figure S4. Optical constants for thick, atomically flat crystals of MoS$_2$ and WS$_2$** (left and right panel, respectively). The crystals were obtained by mechanical exfoliation, using the same method as for the fabrication of heterostructures described in the main text. Here *n* is the refractive index and *k* the extinction coefficient.